\documentclass[onecolumn]{revtex4-2}
\usepackage[utf8]{inputenc}
\usepackage[T1]{fontenc}
\usepackage[]{graphicx}
\usepackage{grffile}
\usepackage{longtable}
\usepackage{wrapfig}
\usepackage{rotating}
\usepackage{amsmath,bm}
\usepackage{textcomp}
\usepackage{amssymb}
\usepackage{capt-of}
\usepackage{hyperref}
\usepackage{natbib}
\usepackage[margin=1in]{geometry}
\setcitestyle{square}
\usepackage[dvipsnames]{xcolor}
\usepackage[breakable,most,many]{tcolorbox}
\usepackage{ulem}
\usepackage{lineno}

\begin{document}

\title{Inferring phase transitions and critical exponents from limited observations with Thermodynamic Maps}


\author{Lukas Herron}
\affiliation{Biophysics Program and Institute for Physical Science and Technology, University of Maryland, College Park, MD, 20742, USA}

\author{Kinjal Mondal}
\affiliation{Biophysics Program and Institute for Physical Science and Technology, University of Maryland, College Park, MD, 20742, USA}


\author{John S. Schneekloth Jr.}
\affiliation{Chemical Biology Laboratory, National Cancer Institute, Frederick, MD 21702, USA}

\author{Pratyush Tiwary*}
\affiliation{Department of Chemistry and Biochemistry and Institute for Physical Science and Technology, University of Maryland, College Park, MD, 20742, USA}

\begin{abstract}
Phase transitions are ubiquitous across life, yet hard to quantify and describe accurately. In this work, we develop an approach for characterizing generic attributes of phase transitions from very limited observations made deep within different phases' domains of stability. Our approach is called Thermodynamic Maps, which combines statistical mechanics and molecular simulations with score-based generative models. Thermodynamic Maps enable learning the temperature dependence of arbitrary thermodynamic observables across a wide range of temperatures. We show its usefulness by calculating phase transition attributes such as melting temperature, temperature-dependent heat capacities, and critical exponents. For instance, we demonstrate the ability of thermodynamic maps to infer the ferromagnetic phase transition of the Ising model, including temperature-dependent heat capacity and critical exponents, despite never having seen samples from the transition region. In addition, we efficiently characterize the temperature-dependent conformational ensemble and compute melting curves of the two RNA systems GCAA tetraloop and HIV-TAR, which are notoriously hard to sample due to glassy-like landscapes.
\end{abstract}

\maketitle

\section{Introduction}


Phase transitions are widely observed in biological, material, and social sciences. Across these disciplines, phase transitions can be defined as the emergence of higher-level, large-scale organization from the coordinated, short-range interactions between many individual constituents. Classical examples include the ferromagnetic to paramagnetic transition, boiling of water, and conformational transitions in biomolecules like proteins and nucleic acids.

Statistical mechanics, and especially the framework of energy landscapes, provides a simple and unifying way of studying phase transitions in these diverse systems \cite{wales-energy-landscapes}. In this work we are specifically interested in phase transitions in systems that stay in equilibrium throughout. For these, the Boltzmann distribution relates the probability of finding a system in a particular microscopic configuration $\mathbf{x}$ to its energy $U(\mathbf{x})$ and the system's inverse temperature $\beta$ as
\begin{equation}
\label{eq:boltzmann}
\mu(\mathbf{x}) = \frac{e^{ -\beta U(\mathbf{x})}}{Z(\beta)} \quad \text{with}
\end{equation}

\begin{equation}
\label{eq:partition-func}
Z(\beta) = \int e^{-\beta U(\mathbf{x})} d\mathbf{x}. 
\end{equation}

$Z(\beta)$ is a normalization constant known as the partition function, whose behavior is often associated with phase transitions. Exploration of the energy landscape is guided by competition between energy and entropy, which is encapsulated by a temperature-dependent free energy $F(\beta)$ which may be computed from the partition function as
\begin{equation}
    \label{eq:free-energy}
    F(\beta) = -\beta^{-1} \ln Z(\beta).
\end{equation}

At a glance, Eqs. \ref{eq:boltzmann}-\ref{eq:free-energy} suggest that the relationship between temperature, energy, microscopic probability, and macroscopic free energy is simple and tractable. It appears as if with these equations, one has the machinery to directly calculate the free energy across temperatures. By doing so for different macroscopic phases one could then obtain various thermodynamic attributes of phase transitions, including transition temperatures and phase diagrams. By calculating appropriate fluctuations, one could directly obtain response functions such as heat capacities and others.

In reality, however, the situation is quite complex. Studying phase transitions and their characteristics computationally is made difficult by Eq. \ref{eq:partition-func}, which requires integrating over a (usually) intractably large number of dimensions. Numerous elegant theoretical and computational schemes have been proposed over the decades to solve this problem. For example, free energy perturbation, Markov chain Monte Carlo (MC) methods,  the replica trick and others \cite{zwanzig-fep,monte-carlo-review,wham,bar,mbar,parisi-replica-trick}.

In this study, we propose a generative Artificial Intelligence (AI) based approach that characterizes phase transitions by learning the temperature dependence of the partition function, and therefore the free energy. Our method, which we call ``Thermodynamic Maps'' (TM) incorporates score-based generative modeling into the framework of free energy perturbation within statistical mechanics \cite{nonequilibrium-learning,ddpm,sgm-sde,ddpm-remd}. The central idea underlying Thermodynamic Maps is that mapping the temperature dependence of ensembles of configurations of a complex system onto the temperature dependence of a simple, idealized system allows for efficient generation of physically realistic samples of the complex system with the correct Boltzmann weights. Within the framework of free energy perturbation, the mapping allows for temperature-dependent free energy estimates. Additionally, Thermodynamic Maps are highly-efficient and can learn from limited data which is not sampled from the global equilibrium distribution.

We demonstrate the applicability of Thermodynamic Maps for three complex systems where we compare against benchmarks from theory, extensive computational studies, and experiments. The first system we consider is the Ising model on a two-dimensional square lattice. With observations made at two temperatures, one deep in the paramagnetic regime and the other deep in the ferromagnetic regime, we are able to correctly infer critical behavior. We then study two different Ribonucleic Acid (RNA) systems: the GCAA tetraloop and HIV-TAR RNA \cite{gcaa-nmr,hiv-nmr}. For both of these, we infer the temperature-dependence of the equilibrium distribution across temperatures using Thermodynamic Maps trained on data generated by short molecular dynamics (MD) simulations. For both RNA systems, we predict the temperature dependence of conformational ensemble and compute melting temperatures in agreement with computational studies and experiment.

Given this demonstration of applicability, we believe that Thermodynamic Maps will be found useful for the characterization of complex phase transitions in diverse systems, especially those with multiple phases. For example, local minima within the energy landscape of biomolecular systems are often biologically and functionally relevant \cite{excited-states}, and while they have been challenging to characterize, we are able to study them with Thermodynamic Maps. The computational efficiency of the learning algorithm and scalability due to not requiring samples from the global equilibrium distribution make Thermodynamic Maps especially suitable for studying large-scale systems exhibiting complex behavior across long timescales.

\section{Method}
\subsection{Targeted Free Energy Perturbation with Machine Learning}

\begin{figure*}
    \centering
    \includegraphics[
    width=\textwidth
    ]{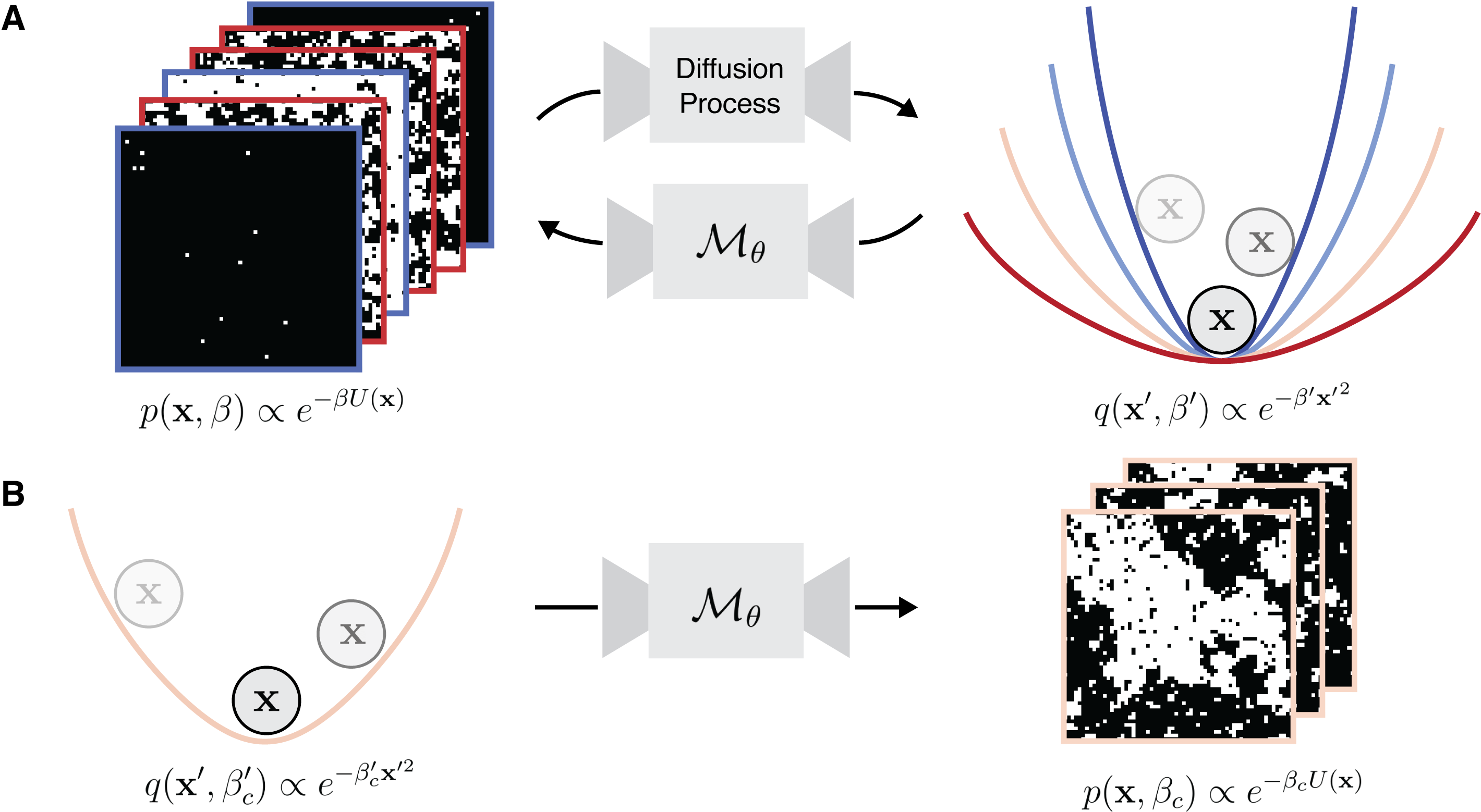}
    \caption{{\bf Illustration of a Thermodynamic Map between systems.} {\bf A} The Thermodynamic Map is parameterized by a diffusion model, denoted as $\mathcal{M}_\theta$, which learns to invert a diffusion process that maps the temperature dependence of samples $\mathbf{x}$ from a complex system, whose equilibrium distribution is $p(\mathbf{x}, \beta)$ onto samples $\mathbf{x}'$ of a simple system whose equilibrium distribution is $q(\mathbf{x}', \beta')$. As pictured, samples data of an Ising model sampled at two temperatures are mapped onto a simple harmonic oscillator. {\bf B} Once learned, the Thermodynamic Map allows samples of the complex system to be generated from the simple prior system at any temperature, even those showing non-trivial behavior.} 
    \label{fig:schematic}%
\end{figure*}



To motivate Thermodynamic Maps, we start with relevant background work. The guiding principle behind all such methods is that one is usually interested in differences of free energies, rather than their absolute values. Such differences can be estimated from ratios of partition functions. Since pioneering work by Zwanzig and Feynman in the 1950s, many frameworks have been developed to do this efficiently \cite{zwanzig-fep,feynman,TI,mbar,bar}.
One such relatively recent framework is that of Targeted Free Energy Perturbation (TFEP) \cite{TFEP}. 

TFEP proposes a potential solution for more efficient estimation of relative free energies, suggesting that estimating the free energy under an invertible mapping
\begin{equation}
    \label{eq:TFEP-mapping}
    \mathcal{M}:\mathbf{x}\rightarrow\mathbf{x}'
\end{equation}
of the configuration space onto itself can enhance the convergence of relative free energy estimates. The motivation behind this approach is that a well-chosen $\mathcal{M}$ can dramatically increase the overlap between states in the configuration space, thus accelerating the convergence of estimates of the relative free energies between them. However, finding such an invertible mapping is difficult in practice due to the complex, high-dimensional distribution of configurations and the requirement that $\mathcal{M}$ preserves dimensionality. 
 
Flow-based generative models have emerged as attractive candidates for this task. These models effectively learn to parameterize high-dimensional, invertible $\mathcal{M}_\theta$ which transforms samples from a simple prior distribution $q(\mathbf{x}')$ into those of a complicated distribution $p(\mathbf{x})$ \cite{NFReview,RealNVP,NeuralODE}. Once the map is learned, samples from $q(\mathbf{x}')$ can be efficiently transformed into samples from $p(\mathbf{x})$ at a lower computational cost than generating samples of $p(\mathbf{x})$ through other means. If $p(\mathbf{x})$ is the image of $q(\mathbf{x}')$ under $\mathcal{M}_\theta$, then probability densities of $q(\mathbf{x}')$ can be transformed into densities of $p(\mathbf{x})$ through

\begin{equation}
    \label{eq:normalization}
    p(\mathbf{x}) = \frac{q(\mathbf{x}')}{| \det J_{\mathcal{M}_\theta}(\mathbf{x}')|},
\end{equation}
where $J_{\mathcal{M}_\theta}$ is the Jacobian of $\mathcal{M}_\theta$. Clearly, using Eq. \ref{eq:normalization} in practice depends on parameterizing $\mathcal{M}_\theta$ which is invertible, has an easy-to-compute Jacobian, and is still expressive enough to transform a simple $q(\mathbf{x}')$ into a complex $p(\mathbf{x})$. Normalizing flows parameterize invertible functions $\mathcal{M}_\theta$ with tractable Jacobians, and are the most widely employed flow-based models for free energy estimation \cite{BG,NF-REMD,NF-MC}. Although normalizing flows are theoretically appealing for free energy estimation, much effort has been spent addressing barriers of application to complex systems.

The main barrier to the wide application of normalizing flows is in training difficulty. The tractable Jacobian comes at the price of reduced expressivity, impeding their ability to map simple priors to complicated target distributions in high dimensions \cite{bbvi}. Recent developments improve robustness by developing more expressive black-box operations with simple Jacobians, or incorporating stochasticity into the flow \cite{smooth-NF,neural-spline,AIS-NF,SNF}. 

Another approach for improving the robustness of normalizing flows is to bring the prior distribution closer to the empirical distribution, thereby simplifying $\mathcal{M}_\theta$. However, this often requires expert knowledge of the system being studied. For example, the Learned Free Energy Perturbation (LFEP) approach enhances free energy estimation of crystalline solids by mapping configurations onto a periodic lattice of particles whose positions are perturbed by Gaussian noise \cite{wirnsberger-TFEP,LFEP}. Free energy differences are first estimated within the tractable prior, and then transformed into free energy estimates of the target distribution using Eq. \ref{eq:normalization}. We take inspiration from this example, where $\mathcal{M}_\theta$ is a mapping from the equilibrium distribution of a physical system onto that of an idealized system.

We extend the applicability of generative modeling to free energy estimation by using score-based models rather than normalizing flows. Score-based models have demonstrated a high degree of expressiveness and robustness in reliably being able to map complex, high-dimensional distributions to trivially simple priors, without requiring a computationally tractable Jacobian \cite{sgm-sde}. Score-based models are generally formulated as pairs of forward and backward stochastic differential equations (SDEs) of the form:

\begin{equation}
\label{eq:sde-fwd-general}
\text{d}\mathbf{x} = -f(\mathbf{x},t)\text{d}t + g(t)\text{d}\mathbf{w}\quad \text{and} \\
\end{equation}
\begin{equation}
\begin{split}
    \label{eq:sde-bck-general}
     \text{d}\mathbf{x} = -\left[f(\mathbf{x},t)  + g(t)^2\nabla_{\mathbf{x}}\log p_t(\mathbf{x})\right]\text{d}t
     + g(t)\text{d}\mathbf{w}.
\end{split}
\end{equation}

 Eq. \ref{eq:sde-fwd-general} defines a diffusion process that relaxes to a prior distribution $q(\mathbf{x}')$ for any distribution of initial conditions $p(\mathbf{x})$. Score-based models exploit the remarkable property that any diffusion process of the form Eq. \ref{eq:sde-fwd-general} can be reversed according to Eq. \ref{eq:sde-bck-general} \cite{Anderson1982}. Together, the stochastic processes of Eqs. \ref{eq:sde-fwd-general} and \ref{eq:sde-bck-general} define a map between distributions $p(\mathbf{x})$ and $q(\mathbf{x}')$ that is exactly invertible in the ensemble limit.
 
However, carrying out the reverse diffusion is not as straightforward as the forward diffusion. The term $\nabla_{\mathbf{x}}\log p_t(\mathbf{x})$, referred to as the score $\mathbf{s}(\mathbf{x},t)$, depends on the initial conditions of Eq. \ref{eq:sde-fwd-general} which generally cannot be expressed in closed-form. Therefore, the score is estimated by a neural network that is trained to match the predicted score to the true score.


Using score-based models, we demonstrate that the temperature dependence of the distribution of structures for a complex system $p(\mathbf{x},\beta)$ can be mapped onto the distribution of structures of a trivial prior system $q(\mathbf{x}',\beta')$. Treating the prior distribution of the score-based model as arising from the dynamics of a physical system allows us to infer the temperature dependence of free energies, extending TFEP to the multi-ensemble case.

\subsection{Thermodynamic Maps}
\label{sec:TM}

To extend TFEP to multi-ensemble thermodynamics, we augment our coordinates $\mathbf{x}\in \mathbb{R}^{d}$ with auxiliary inverse temperature-type variables $\bm{\beta}\in \mathbb{R}^{d}$. Together, the coordinates and temperatures form a state vector $(\mathbf{x},\bm{\beta})^\top \in \mathbb{R}^{2d}$, and the thermodynamic map $\mathcal{M}_\theta$ is parameterized in the joint $\mathbf{x}$-$\bm{\beta}$-space as a score-based model. For a discussion on the role of $\bm{\beta}$ in learning thermodynamic relationships, refer to Appendix \ref{appendix:fluctuations}.

We express $\mathcal{M}_\theta$ as a pair of forwards and backward SDEs, as is standard for score-based models:

\begin{equation}
\label{eq:sde-fwd}
\begin{pmatrix}
    \text{d}\mathbf{x}\\
    \text{d}\bm{\beta}^{-1}
\end{pmatrix}
 = -\frac{1}{2}\sigma(t)
\begin{pmatrix}
    \mathbf{x}\\
    \bm{\beta}^{-1}
\end{pmatrix}
\text{d}t + \sqrt{\sigma(t)}
\begin{pmatrix}
    \sqrt{\bm{\beta}_0^{-1}}\\
    \mathbf{1}
\end{pmatrix}\text{d}\mathbf{w}\quad \text{and}
\end{equation}


\begin{equation}
\label{eq:sde-bck}
\begin{pmatrix}
    \text{d}\mathbf{x}\\
    \text{d}\bm{\beta}^{-1}
\end{pmatrix} = -\frac{1}{2} \sigma(t) \left[
\begin{pmatrix}
    \mathbf{x}\\
    \bm{\beta}^{-1}
\end{pmatrix} + 
\begin{pmatrix}
    \mathbf{s}_\theta(\mathbf{x},t)\\
    \mathbf{s}_\theta(\bm{\beta}^{-1},t)
\end{pmatrix}\right]\text{d}t + \sqrt{\sigma(t)}
\begin{pmatrix}
    \sqrt{\bm{\beta}_0^{-1}}\\
    \mathbf{1}
\end{pmatrix}\text{d}\mathbf{w},
\end{equation}
where $\mathbf{w}$ is isotropic, unit Gaussian noise. Eq. \ref{eq:sde-fwd} governs the forward diffusion of the coordinates and inverse temperatures, with monotonically increasing $\sigma(t)$ chosen so that $\sigma(0)=0$ and $\sigma(T)=1$. Starting from initial conditions $\mathbf{x}_0$ and $\bm{\beta}_0$ at $t=0$, the coordinates and temperatures converge to $\mathcal{N}(\mathbf{0}, \bm{\beta}_0^{-1})$ and $\mathcal{N}(\mathbf{0}, \bm{1})$ respectively at $t=T$. 


The score-based model defined by Eqs \ref{eq:sde-fwd} and \ref{eq:sde-bck} are motivated by treating the prior distribution of the coordinates as the equilibrium distribution arising from the dynamics of some physical system. If the prior is a normal distribution, then the corresponding physical system is a harmonic oscillator, and the temperature $\bm{\beta}_0$ corresponds to the oscillator's temperature.  Generating samples at a particular temperature amounts to sampling from the prior at the corresponding temperature in Eq. \ref{eq:sde-bck}. Optionally, one may hold the temperatures fixed over the course of the generative process so that $\mathcal{M}_\theta$ is restricted to the configuration space as required by TFEP. Empirically we find that diffusion in the joint $\mathbf{x}$-$\bm{\beta}$-space yields similar results to the conditional generation in \cite{ddpm-remd}. More details on training and inference are provided in Appendix \ref{appendix:TM}



\subsection{Critical Behavior of the Ising Model}

We apply Thermodynamic Maps to the most widely studied formulation of the classical Ising model as a system of interacting spins arranged on a two-dimensional square lattice, without the presence of an external magnetic field. Arguably, this model serves as a prototypical example of a simple system with a complex phase transition. Each spin can have one of two states with value $\sigma=+1$ or $\sigma=-1$. The spins interact with their nearest neighbors through an interaction term $J$ and Hamiltonian given by 
\begin{equation}
    \mathcal{H} = J\sum_{\langle i,j\rangle } \sigma_i \sigma_j,
\end{equation}
where $\langle i,j\rangle$ denotes nearest-neighbor pairs. When $J<0$ the model exhibits ferromagnetic behavior in its ground state. Going forward, we set $J=-1$ without loss of generality.
\begin{figure*}
    \centering
    \includegraphics[
    width=\textwidth
    ]{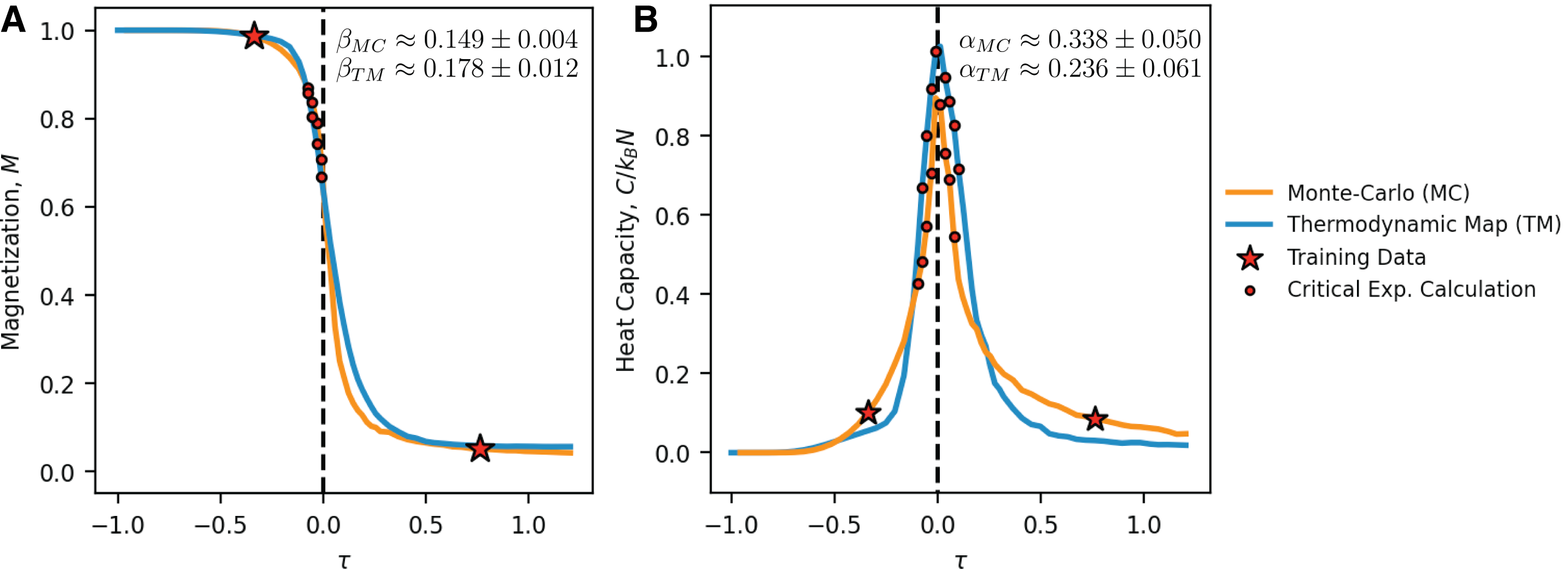}
    \caption{{\bf Inferring the phase transition of the 2D Ising model from limited sampling.} {\bf A} The magnetization is plotted for samples of a $32\times32$ square Ising model generated through MC sampling (orange) and the thermodynamic map (blue). The thermodynamic map predicts change in magnetization at $T_c$ when trained on samples generated at $T=1.5$ and $T=4$ (red stars). {\bf B} The heat capacity of samples generated from MC sampling (orange) and the thermodynamic map (blue) is plotted. The thermodynamic map correctly infers the divergence in the heat capacity, numerically computed for the red dots, when trained on the same samples as panel A (red stars).}
    \label{fig:ising-model-mag}%
\end{figure*}

As the temperature increases, an Ising model in two or higher dimensions transitions from an ordered magnetic phase to a disordered paramagnetic phase. For our set-up of a two-dimensional Ising model on a square lattice with $J=-1$, this critical temperature is known to be $T_c\approx2.27$ \cite{onsager-solution}. The two phases can be distinguished from each other with the magnetization order parameter, $M$, defined as the absolute value of the average of the spins. Above $T_c$, the spins are equally likely to be +1 or -1, regardless of their neighbors, resulting in a net magnetization of zero. Well below $T_c$, all spins in the lattice align, leading to a magnetization of 1. Divergences are a signature of critical behavior, and the  magnetization $M$ and heat capacity $C$ of the Ising model diverge near the critical temperature as:
\begin{equation}
    M \sim |\tau|^{\beta} \quad \mathrm{and} \quad C \sim |\tau|^{-\alpha} \quad \mathrm{where} \quad \tau = \frac{T-T_c}{T_c},
\end{equation}
and with critical exponents $\alpha=0$ and $\beta=0.125$ {(not to be confused with the inverse temperature variables $\bm{\beta}$), which can be derived analytically in the thermodynamic limit \cite{ising-crit-exp}. For systems which are not solvable the critical exponents must be measured numerically, which is often done through MC sampling, wherein proposals for spin flips are generated and their acceptance or rejection is determined based on the detailed balance condition. Near the critical temperature, the presence of long-range correlations causes MC dynamics to slow down exponentially, leading to difficulty in sampling the phase transition \cite{wolff-critical-slow-down}.

We investigate the ability of Thermodynamic Maps to infer such critical behaviors by generating configurations of an Ising model through MC sampling across temperatures, and using the data from two temperatures asymmetrically spaced about $T_c$ to train a TM. We then generate configurations at all temperatures using the TM, and compare the behavior of $M$ and $C$ between the MC and TM-generated samples, as shown in Figure \ref{fig:ising-model-mag}. The TM infers the correct value of $T_c$ ($\tau=0$) even on the basis of limited, deliberately misleading training data, and generates samples with divergences in $M$ and $C$. 

Figure \ref{fig:ising-model-mag}A shows the behavior of the magnetization for MC and TM-generated samples across temperatures, with critical exponents measured as $\beta_{MC} \approx 0.149 \pm 0.004$ and $\beta_{TM} \approx 0.178 \pm 0.012$. Similarly, Figure \ref{fig:ising-model-mag}B shows a divergence in the heat capacity at $T_{MC}\approx2.25$ and $T_{TM}\approx2.30$, with critical exponents $\alpha_{MC} \approx 0.338 \pm 0.050$ and $\alpha_{TM} \approx 0.236 \pm 0.061$. For both exponents, the MC and TM exponents are in agreement with each other, although far from the ideal value due to finite size effects.

These results highlight the ability of our model to infer the non-trivial thermodynamic behavior of phase transitions without being shown samples from the transition region; the physical meaning of temperature in the prior system has been transferred to the complex system, complete with the statistics properties associated with critical behavior. The correct prediction of $T_m$, even with asymmetrically spaced training temperatures, indicates our model has learned the physics of the Ising model, not merely the distribution of structures at each temperature.



\subsection{Exploring RNA Conformational Landscapes}
\label{sec:RNA}

\begin{figure*}%
    \centering
    \includegraphics[
    width=\textwidth
    ]{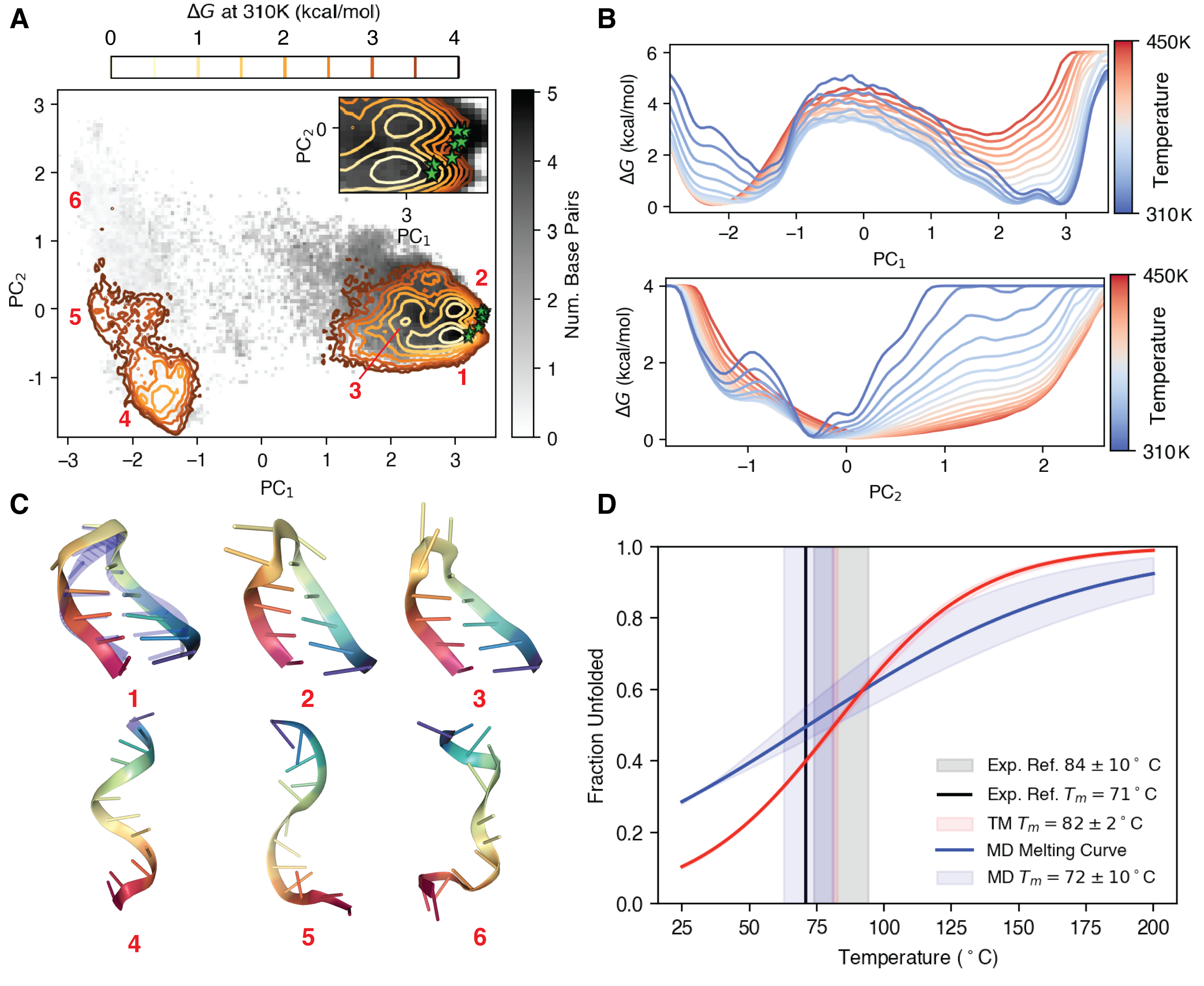}
    \caption{{\bf GCAA Tetraloop Conformational Landscape.} {\bf A} Joint distribution of the first two principal components of G-vectors for GCAA. Contours representing TM-generated samples at 310K are overlaid on MD samples shaded by the number of base pairs, with selected regions of the conformational landscape annotated as 1-6. The 10 NMR conformers reported in the Protein Data Bank (PDB: 1ZIH) are depicted as green stars. {\bf B} Temperature-dependent free energy profiles along the first two principal components shown in panel A. Colors indicate temperatures ranging from 310K to 450K at 10K intervals. {\bf C} Representative structures sampled from labeled minima in panel A, with a representative NMR conformer shown in translucent blue. The structures are colored from blue to red from the 5' to 3' end. {\bf D} A fraction folded curve is obtained by reweighting the MD data in panel A with the TM conformational ensemble for each temperature in panel B, and fitting a two-state model. Uncertainties are computed from the last three iterations of TM-aMD. The cutoff for folded and unfolded states is three base pairs to match our reference MD study \cite{DE-Shaw-RNA-ff}.}
    \label{fig:gcaa}%
\end{figure*}

To show the broad applicability of Thermodynamic Maps, we now study conformational transitions and melting in two different ribonucleic acid (RNA) systems. Studying atomic-resolution conformational ensembles of RNAs through molecular dynamics simulations has proved crucial for understanding RNA structural dynamics, yet remains challenging due to the disordered, glassy nature of RNA energy landscapes \cite{DE-Shaw-RNA-ff,bussi-ff-deficiency}.

Increasing evidence points towards some RNAs having glassy energy landscapes where multiple minima are separated by high barriers \cite{kremer-polymer,parisi-RNA,wales-RNA}. The ruggedness of the landscape results in many competing degrees of freedom and no clear-cut separation of timescales within the dynamics \cite{hashimi-glass,wales-RNA,parisi-RNA}. The most striking feature of glassy energy landscapes as they relate to polymers is the presence of conformational heterogeneity at equilibrium \cite{parisi-RNA, kremer-ML}.  While the conformational landscape of proteins is often dominated by a single well-defined fold (energy minimum), the RNA conformational landscape may not dominated by a single structure \cite{hashimi-invisible,hashimi-nmr-dock}. This difference is analogous to the single magnetized phase of an Ising model and the many phases of long-ranged spin glasses \cite{parisi-replica-trick, ANNNI,hopfield-network,amit-spin-glass}.  Since multiple members of the ensemble contribute substantially to the free energy and other thermodynamic observables, RNAs are best described as a weighted ensemble of conformers \cite{how-to-think,hashimi-nmr-dock}. 

Exploring energy landscapes by biasing dynamics along a small number of slow degrees of freedom has proven successful for exploring the conformational landscape of proteins, but the lack of timescale separation and ensemble nature of the RNA conformational landscape violates fundamental assumptions of dimensionality made in biasing methods \cite{metadynamics-review}. On the other hand, Thermodynamic Maps have the advantage of being able to learn the conformational landscape directly in the high-dimensional configuration space.

We train Thermodynamics Maps on limited information generated by bioinformatic approaches and multi-ensemble molecular dynamics simulations to efficiently characterize RNA conformational ensembles. Our starting point is the physics and knowledge-based potential that is central to Rosetta to generate a putative conformational ensemble \cite{rosetta}. These structures serve as the starting point to explore a more realistic energy landscape through all-atom, explicit solvent molecular dynamics performed over a range of temperatures. Between rounds of molecular dynamics simulation, the global equilibrium distribution is inferred using Thermodynamic Maps. Initial conditions are re-sampled from the inferred equilibrium distribution according to a general RNA order parameter. Here on, we refer to this protocol as Thermodynamic Map-accelerated Molecular Dynamics (TM-aMD).

Although agreement between the input equilibrium distribution and the output from the Thermodynamic Map is a necessary, though not sufficient condition that the true equilibrium distribution has been attained, we leave rigorously addressing convergence to equilibrium to future work. Here we take solace in the numerical results for the two challenging test systems, described next. 

\subsubsection{GCAA Tetraloop}

With a shift in perspective towards viewing RNAs as dynamic entities, there has been interest in studying the variation in dynamics between different tetraloops. Studying combining computation and experiment have demonstrated that even so-called simple tetraloops can exhibit rich dynamics \cite{kresten-jacs}. We study the GCAA tetraloop, a well-studied model system, enabling us to compare the equilibrium distribution generated by our model with extensive molecular dynamics simulations and experimental data \cite{debenedetti-rna,DE-Shaw-RNA-ff,gcaa-melt}. The GCAA Tetraloop is a small, highly-stable, 12-nucleotide RNA sequence that adopts a hairpin structure consisting of an eight-nucleotide helix and a four-nucleotide loop (PDB: 1ZIH) \cite{gcaa-nmr}. Consistent with an ensemble perspective, the variable arrangement of nucleotides in the loop gives rise to alternative conformations, which we investigate with TM-aMD.

The Thermodynamic Map learns to generate RNA structures represented as $G$-vectors, which is an internal coordinate system for RNAs that effectively clusters distinct folded states \cite{gvecs-bussi}. The principal components of $G$-vectors have been shown to be a convenient visualization of RNA structural diversity, which we use to guide adaptive sampling. Further information on $G$-vectors, along with details of TM-aMD can be found in Appendix \ref{appendix:TM-aMD}.

Figure \ref{fig:gcaa} summarizes the result of nine iterations of our enhanced sampling procedure, with a total of 50$\mu$s of simulation. Although we performed extensive MD simulations reaching long timescales, we still used two orders of magnitude less compute compared to our reference millisecond replica exchange simulation, even with sub-optimal scheduling of TM learning with MD simulation  \cite{DE-Shaw-RNA-ff}. Supplementary Figure \ref{fig:convergence}A suggests that TM-aMD would benefit from more frequent reseeding of simulations.


In Figure \ref{fig:gcaa}A, we present the projection of the learned equilibrium distribution onto the first two principal components of the $G$-vectors. The contours represent the free energy landscape inferred by the thermodynamic map, while the shaded regions represent the distribution of structures observed in the simulation, shaded by the number of base pairs. The 10 conformers reported in the Protein Data Bank (PDB), represented as green stars, lie within the most dominant TM-predicted cluster. Figure \ref{fig:gcaa}B shows the learned, temperature-dependent free energy profiles along each of the principal components in Figure \ref{fig:gcaa}A. The first principal component clearly shows the reweighting of the folded and unfolded states with temperature. Figure \ref{fig:gcaa}C shows the structures associated with the minima in Figure \ref{fig:gcaa}A. The first three conformations are abundant at 310K and are consistent with MD studies \cite{DE-Shaw-RNA-ff,debenedetti-rna} and experiment (an example NMR conformer is depicted in blue). The last three represent unfolded states that are stabilized by base stacking and are weakly present in the 310K ensemble. The first principal component corresponds to the folding-unfolding transition, while the second captures the conformational heterogeneity of the loop region.

Figure \ref{fig:gcaa}D displays the melting curve of the GCAA tetraloop computed from the last three rounds of MD simulation, with the melting curve of the learned equilibrium distribution. Both curves exhibit agreement, matching the range of melting temperatures predicted from a lengthy replica exchange simulation \cite{DE-Shaw-RNA-ff}. 

\subsubsection{HIV-TAR RNA}

\begin{figure*}%
    \centering
    \includegraphics[
    width=\textwidth
    ]{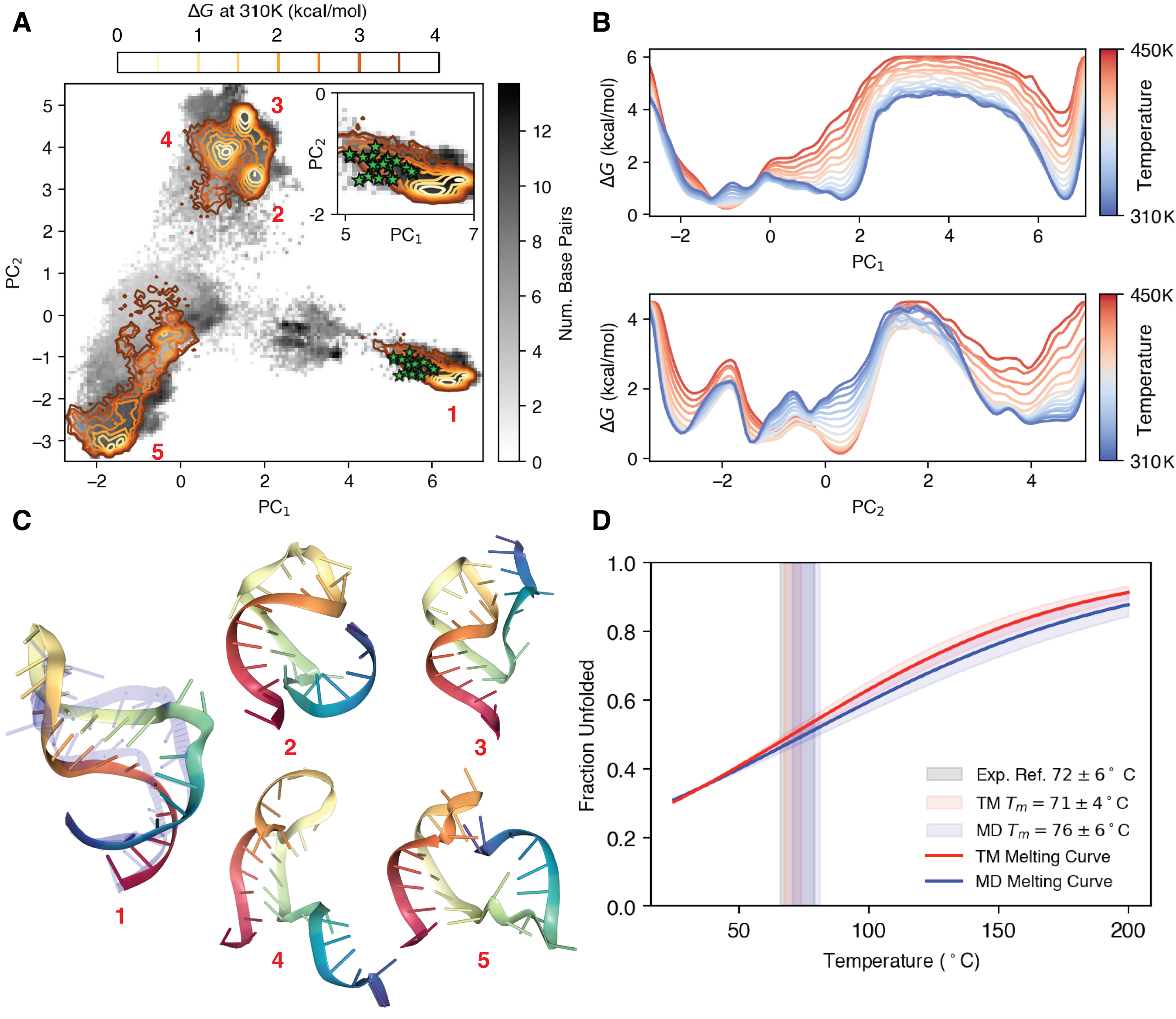}
    \caption{{\bf HIV-TAR RNA Conformational Landscape.} {\bf A} Joint distribution of the first two principal components of G-vectors for HIV-TAR. Contours representing TM-generated samples at 310K are overlaid on MD samples shaded by the number of base pairs, with basins of the TM landscape labeled 1-5. The 20 reported NMR conformers in the Protein Data Bank (PDB: 1ANR) are depicted as green stars. {\bf B} Temperature-dependent free energy profiles along the first two principal components are shown in panel A. Colors indicate temperatures ranging from 310K to 450K at 10K intervals. {\bf C} Representative structures sampled from labeled clusters in panel A, with a representative NMR conformer shown in translucent blue. The structures are colored from blue at the 5' end to red at the 3' end.  {\bf D} A fraction folded curve is obtained by reweighting the MD data in panel A with the TM conformational ensemble for each temperature in panel B, and fitting a two-state model. Uncertainties are computed from iterations of the TM-aMD algorithm showing agreement between MD and TM-predicted melting temperatures (see Appendix \ref{appendix:convergence}). The cutoff for folded and unfolded states is nine base pairs, as determined by the NMR conformers).}
    \label{fig:hiv}%
\end{figure*}

The HIV-TAR RNA is an extensively studied, 29 nucleotide RNA hairpin from the HIV genome that displays rich conformational diversity. The secondary structure consists of lower and upper helices separated by a three-nucleotide bulge with an apical loop closing the upper helix (PDB: 1ANR) \cite{hiv-nmr}. The disordered loop and bulge regions mediate interactions with proteins and small molecules \cite{hiv-tar-binding}. We investigate the conformational landscape of the HIV-TAR RNA using six iterations of TM-aMD, requiring a total of 70$\mu$s of simulation time. We infer relative free energies between the dominant conformer observed through NMR spectroscopy and alternative conformers, which are unattainable through MD simulation alone.


The conformational landscape predicted by the learned TMs, projected along the principal components of the $G$-vectors is shown in Figure \ref{fig:hiv}A. Structures corresponding to the NMR ensemble (green stars) are well-separated from other misfolded structures within the two-dimensional projection. The temperature dependence of the free energy along each principal component is shown in Figure \ref{fig:hiv}B, where we find that the free energy barrier separating the NMR conformers from unfolded states reaches a height of $5$kcal/mol, which is approximately the free energy of breaking a canonical base pair. Figure \ref{fig:hiv}C depicts representative samples from each cluster.  The first cluster shows agreement with the NMR conformers, with the first reported conformer shown in translucent blue. The other clusters correspond to varied secondary structure motifs. Generally, across the conformational landscape, we find that the contribution of folded states to the conformational ensemble diminishes as temperature increases. This can be clearly shown by computing a fraction folded curve (Figure \ref{fig:hiv}D), which shows agreement between melting temperatures of the MD and TM-derived ensembles, and the experimental melting temperature \cite{hiv-experiment}.

Our results support the idea that RNAs evolve over a rugged free energy landscape punctuated by many long-lived states \cite{dill-chen}. We find that the native state is indeed the most stable at physiological temperatures ($310$K), but misfolded states consisting of secondary structure elements still have substantial contributions to the ensemble. Minima of the energy landscape corresponding to states 2-5 in Figure \ref{fig:hiv} differ from the NMR conformers (state 1) by less than $1$kcal/mol. Overall, our findings are in agreement with theoretical studies of RNA folding pointing towards a rugged energy landscape \cite{dill-chen,wales-RNA}.


Clearly, our findings are dependent on the accuracy of the simulation force field, which are notoriously inaccurate for RNAs, and our algorithm is incapable of reaching experimental folding timescales. Despite this, the fraction folded curve is in agreement with experimental results. As such, TMs may be useful in the context of forcefield development, where a major barrier to improved forcefields is the difficulty in attaining equilibrium through MD simulation, especially for systems that have many metastable states.

\section{Discussion}

We have demonstrated that Thermodynamic Maps are capable of accurately inferring the non-trivial behavior of the free energy with temperature from limited data for the two-dimensional square Ising model. Additionally, we also show how Thermodynamic Maps may be integrated with molecular dynamics simulation to enhance sampling. In this section, we outline prospective, potential applications for Thermodynamic Maps.


Our RNA results indicate that Thermodynamic Maps are well suited for inferring the equilibrium distribution from swarms of short, independent simulations. We show that TM-aMD is capable of exploring the conformational landscape orders of magnitude faster than REMD, while still inferring global equilibrium. Still, TM-accelerated MD stands to benefit greatly from optimization. And since TMs allow for truly parallel simulations across arbitrary temperatures, TMs are poised to make efficient use of distributed computing resources \cite{folding-at-home}. 

As pointed out in Section \ref{sec:RNA}, we believe that our implementation of TM-aMD is hindered by the scheme used to adaptively re-sample initial states. The one we employ aims for agreement between the TM and MD-generated distributions. However, in practice, it is more productive to balance thermodynamic accuracy with directed sampling of diverse conformations. Methods such as FAST and REAP quantify this trade-off as a reward function which is to be optimized by the choice of adaptive sampling strategy \cite{FAST,REAP}.




The large number of observed metastable states for small RNAs such as HIV-TAR indicate that a thorough characterization of RNA structure at equilibrium will require reaching simulation timescales that are longer than can be reached through brute-force computing alone. We sidestep the timescale problem by starting our exploration from a putative ensemble generated by the coarse-grained Rosetta potential. We leave the more computationally intense task of truly {\it de novo} folding of RNAs through MD simulations to future work.

Overall, we have presented Thermodynamic Maps (TMs) as an exciting way of integrating score-based generative modelling with the theoretical framework of statistical mechanics. We have demonstrated the capability of Thermodynamic Maps to learn physics by predicting the critical behavior associated with the Ising Model from limited, deliberately misleading sampling. We also use Thermodynamic Maps to accelerate Molecular Dynamics (TM-aMD) and show that TM-aMD efficiently infers the equilibrium distribution of two model RNA systems using a fraction of the computational resources required by Replica-Exchange simulations. Based on these results, we believe that Thermodynamic Maps are suitable for widespread use and have great potential for further theoretical development and computational optimization.

\section{Acknowledgements}
This research was supported by the Intramural Research Program of the National Institutes of Health, National Cancer Institute (NCI), Center for Cancer Research, Project BC011585 09 (J.S.S.) L.H. was supported by the National Science
Foundation, Grant No. CHE-2044165. J.S.S. and P.T. thank the NCI-UMD Partnership for Integrative Cancer Research. P.T. was an Alfred P. Sloan Foundation fellow during preparation of this manuscript. We thank Zaratan and XSEDE (project CHE180027P) for computational resources. L.H. would like to thank Yihang Wang and Disha Sanwal for insightful discussions, and Eric Beyerle and Suemin Lee for critical reading of our manuscript.

\appendix

\section{Thermodynamic Maps}
\label{appendix:TM}

\subsection{Network Architecture and Parameters}

Our Thermodynamic Map is parameterized as a score-based model using a U-net \cite{unet} with four Resnet downsampling blocks, followed by an attention layer, and four Resnet upsampling blocks. For all experiments, we use a learning rate of $10^{-3}$ with the Adam optimizer \cite{adam}. We also employ self-conditioning to accelerate training convergence \cite{ddpm-self-conditioning}. The Thermodynamic Map is trained for 250 epochs with 100 diffusion timesteps for all studies performed. We use the polynomial noise schedule of Hoogeboom et. al. \cite{equivariant-diffusion-model}

\subsection{Training Procedure}

As score-based models, Thermodynamic Maps are trained to estimate the score $\mathbf{s}(\mathbf{x},t)=\nabla_{\mathbf{x}}\log p_t(\mathbf{x})$ of Eq. \ref{eq:sde-fwd} so that the forward diffusion in Eq. \ref{eq:sde-fwd} can be reversed according to Eq. \ref{eq:sde-bck}. With the goal of score matching laid out, the training algorithm is as follows.

To generate examples of the score, randomly selected samples $\mathbf{u}_0 = (\mathbf{x}_0,\bm{\beta}_0)^\top$ are corrupted by evaluating the forward diffusion (Eq. \ref{eq:sde-fwd}) up to a uniformly sampled timestep $t \in \mathcal{U}(0,T)$. In practice, the forward diffusion is approximated by a Markov chain with transitions taking the form
\begin{equation}
    \label{eq:markov-chain}
    \mathbf{u}_i = \sqrt{1-\sigma_i} \mathbf{u}_{i-1} + \sqrt{\sigma_i} \mathbf{z} ,
\end{equation}
where $\mathbf{z}$ is sampled from $\mathcal{N}(\mathbf{0}, \bm\beta_0^{-1})$ for the dimensions of $\mathbf{u}$ corresponding to $\mathbf{x}$, and $\mathcal{N}(\mathbf{0}, \mathbf{1})$ for the dimensions corresponding to $\bm{\beta}$. The continuous noise schedule $\sigma(t)$ is discretized as $\{\sigma_i\}_{i=1}^N$, where $N$ is the number of diffusion timesteps. 

The Markov chain defined by Eq. \ref{eq:markov-chain} linearly interpolates between data $\mathbf{u}_0$ and noise $\mathbf{z}$. As such, Eq. \ref{eq:markov-chain} can be evaluated up to any $\mathbf{u}_i$ starting from $\mathbf{u}_0$ in a single step by
\begin{equation}
    \mathbf{u}_i = \sqrt{1-\alpha_i}\mathbf{u}_0 + \sqrt{\alpha_i}\mathbf{z},
\end{equation}
where $\alpha$ is a cumulative noise schedule defined as
\begin{equation}
    \label{eq:marginal}
    \alpha(t) = e^{-\frac{1}{2}\int_0^t \sigma(t')dt'}
\end{equation}
and has been discretized the same as $\sigma(t)$. In practice, we directly specify $\alpha(t)$ as a monotonically increasing function on $[0,T]$ rather than computing the cumulative noise schedule from $\sigma(t)$.

The U-net backbone of the model, denoted as $f_\theta$ takes noisy samples $\mathbf{u}_i$ as input, and is trained to return denoised samples by minimizing the loss
\begin{equation}
    \label{eq:loss}
    \mathcal{L} = \mathbb{E}_{\mathbf{s}_0} \mathbb{E}_{t\sim\mathcal{U}(0,T)} \left[ ||\mathbf{u}_0 - f_\theta \left(\mathbf{u}_i, t\right)||^2 \right].
\end{equation}
Eq. \ref{eq:loss} is equivalent to score matching and has the form of the simple loss defined in \cite{ddpm}. Predicting $\mathbf{u}_0$ from $\mathbf{u}_i$ enables one to estimate the noise added using Eq. \ref{eq:marginal}, which in turn is equivalent to the score up to a multiplicative factor determined by the cumulative noise schedule \cite{three-interpretations}.

The denoised observations $f_\theta(\mathbf{u}_i,t)$ can be used to evaluate the reverse diffusion process through
\begin{equation}
    \mathbf{u}_{i-1} = \sqrt{\alpha_{i-1}} f_\theta (\mathbf{u}_i, t) + \sqrt{1 - \alpha_{i-1} - \eta^2_i} \left( \frac{\mathbf{u}_i - \sqrt{1-\alpha_i}f_\theta(\mathbf{u}_i,t)}{\sqrt{\alpha_i}} \right) + \eta_i^2 \mathbf{z},
\end{equation}
which is equivalent to the generative process of Denoising Diffusion Implicit Models (DDIM) \cite{ddim}. DDIM treats $\eta_i^2$, the variance of the noise in the reverse diffusion, as a hyperparameter that affects the generative process. When $\eta_i = \sqrt{(1-\alpha_i)/(1-\alpha_i)} \sqrt{1-\alpha_i/\alpha_{i-1}}$ the generative process approximates Eq. \ref{eq:sde-bck}. However, when $\eta_i = 0$, the generative process approximates the so-called probability flow ordinary differential equation which is equivalent to Eq. \ref{eq:sde-bck} at the ensemble level of description. The generative process is evaluated with $\eta_i=0$ for all studies performed.

\section{Thermodynamic Map-accelerated Molecular Dynamics (TM-aMD)}
\label{appendix:TM-aMD}

\subsection{G-vectors}
$G$-vectors are an internal coordinate system that represents RNA structures as an orientational, equivariant contact map between all pairs of nucleobases \cite{gvecs-bussi}. The $G$-vector between nucleobases $i$ and $j$ is computed from the distance between the centers of the six-membered rings of the bases $\mathbf{r}_{ij}$ as
\begin{gather}
    \mathbf{G}_{ij}(\mathbf{r}_{ij}) = 
    \begin{pmatrix}
    \sin (\gamma r_{ij})\frac{r_{ij,x}}{r_{ij}} & \\
    \sin (\gamma r_{ij})\frac{r_{ij,y}}{r_{ij}} & \\
    \sin (\gamma r_{ij})\frac{r_{ij,z}}{r_{ij}} &  \\
    1 + \cos (\gamma r_{ij})
    \end{pmatrix}
    \times \frac{\Theta (r_{ij} - r_c)}{\gamma},
\end{gather}
where $r_{ij}$ is the distance between bases $i$ and $j$, $\Theta$ is the Heaviside step function, and $r_c$ is a cutoff of $2.4\AA$, and $\gamma=\pi/r_c$. The internal representation of an RNA with $N$ bases then takes the form of a $N \times N \times 4$ dimensional tensor, making the $G$-vector representation convenient for use with convolutional networks.

\subsection{Rosetta, MD simulations, and Adaptive Sampling}

Initially, 10,000 structures are generated through MC sampling on the Rosetta potential \cite{rosetta}. $G$-vectors are computed for all generated structures, and candidate structures are sampled via regular space clustering on the first two principal components of the $G$-vectors. Before clustering, we compute the distribution energies of the generated structures in vacuum using the MD force field, and we discard any structures with energy greater than the mean plus two standard deviations.

These initial structures are then simulated using the OpenMM simulation engine \cite{openmm} with the D.E. Shaw RNA force field \cite{DE-Shaw-RNA-ff} at $1$M KCl concentration and $1$ atm with the TIP-4P water model \cite{tip4p}. Simulation temperatures are uniformly spaced at $10$K intervals between $310$K and $450$K, and are assigned to Rosetta structures according to increasing energy. 

Structures are equilibrated for 10ns before the first production simulation of 50ns. Between rounds of adaptive sampling, structures are equilibrated for 1ns. Simulation structures are saved at $1$ns intervals during the production simulation, and these structures are represented as $G$-vectors to form the training data for the Thermodynamic Map.

Beyond the initial round of simulation beginning from Rosetta structures, simulations are launched from new structures which are adaptively sampled via regular space clustering on 100000 samples generated by the thermodynamic map for each temperature on the simulation ladder (uniformly spaced at $10$K intervals between $310$K and $450$K). Only samples with free energy less than $2$kcal/mol relative to the most abundant conformation are considered for adaptive sampling at each temperature.

\subsection{Fluctuation Mapping}
\label{appendix:fluctuations}
Thermodynamic maps need not rely on the assumption of data sampled from a global equilibrium distribution. The statistics of data sampled from a local equilibrium distribution (related to the global equilibrium distribution by a shift in the free energy) will not represent the statistics of the global equilibrium distribution \cite{tram}. In this case, using the bath temperature for the $\bm{\beta}^{-1}$ variables will be inaccurate.

For MD simulations, we instead use the mean square fluctuation along each coordinate for $\bm{\beta}^{-1}$ in place of the bath temperature. With this modification, fluctuations of the complex simulated system match the fluctuations of the prior system. Within the prior system, the mean square fluctuation is equivalent to the bath temperature \cite{landau-lifshitz}, so to make predictions at a simulation temperature, we fit a linear relationship between the simulation bath temperature and the mean square fluctuations of the prior. Figures \ref{fig:gcaa} and \ref{fig:hiv} demonstrate that fluctuation mapping produces good agreement with simulation and experiment, and we leave a more rigorous, detailed investigation of fluctuation mapping to future work.

\subsection{Ensemble Weighted Observables}

Consistent with an ensemble perspective of biomolecular structure, we compute ensemble-weighted fraction folded curves for the studied RNA systems (Figures \ref{fig:gcaa}D and \ref{fig:hiv}D). In principle, any observable $A(\mathbf{x})$ computed from microscopic coordinates $\mathbf{x}$ can be weighted by the conformational ensemble at temperature $\beta^{-1}$, denoted as $p(\mathbf{x},\beta)$, through 
\begin{equation}
    \label{eq:ensemble-weighting}
    \langle A(\mathbf{x},\beta)\rangle_p = \langle A(\mathbf{x})|p(\mathbf{x},\beta) \rangle,
\end{equation}
where $\langle \cdot | \cdot \rangle$ denotes the inner product. Estimating $p(\mathbf{x},\beta)$ in Eq. \ref{eq:ensemble-weighting} through binning (as done in Section \ref{sec:RNA}) is usually intractable for high-dimensional $\mathbf{x}$. For this reason, it is often more effective to describe the system in terms of some lower-dimensional collective variables $\xi = f(\mathbf{x})$. For example, the ensemble-weighted number of base pairs in Section \ref{sec:RNA} is computed along the first two principal components of the $G$-vectors for RNA structures, with measurements of $A(\xi)$ gathered from MD simulations, and $p(\xi,\beta)$ generated by the Thermodynamic Map. 

\subsection{Convergence to Equilibrium}
\label{appendix:convergence}
A central unanswered question is how to determine convergence to the equilibrium distribution for TM-aMD. Other methods, such as those based on Markov state models, rely on enforcing the detailed balance condition for transitions between states \cite{msm}. However, the samples produced by Thermodynamic Maps are statistically independent, and checking detailed balance in this manner requires correlated samples. Rather than checking the detailed balance conditions, we argue that if the Thermodynamic Map has learned the equilibrium distribution, then the MD and TM-generated distributions should be in agreement and should not change between iterations. As stated in the main text, this condition is necessary, although not completely sufficient to identify equilibrium. For simplicity, instead of monitoring agreement between the MD and TM distributions, we monitor agreement between MD and TM predictions of observables. 

More concretely, for the RNA systems studied we monitor agreement between melting temperature predictions, as shown in Supplementary Figure \ref{fig:convergence}. For the GCAA tetraloop (Supplementary Figure \ref{fig:convergence}A), we observe disagreement between MD and TM-predicted melting temperatures until $\sim45\mu$s of MD simulation, after which the melting temperatures show agreement and are unchanging between iterations. The final three iterations satisfy our heuristic equilibrium condition and are used to compute the melting temperature reported in the main text (Figure \ref{fig:gcaa}). On the other hand, the melting temperature predictions for HIV-TAR do not show the same convergence. Supplementary Figure \ref{fig:convergence}B shows that for four of the six iterations, the MD and TM-predicted melting temperatures are in agreement, but are clearly not converged. Nonetheless, we use the four iterations showing agreement (3,5,6,7) to compute the melting temperature reported in the main text.

Another striking feature of the convergence plots (Supplementary Figure \ref{fig:convergence}) is the instability of melting temperature predictions for simpler GCAA tetraloop, which occur over a range of $500^\circ$C. Examining the TM-generated conformational landscape at each iteration explains these highly variable predictions. Supplementary Figure \ref{fig:gcaa-contours} shows the TM-predicted conformational landscape at each iteration of Supplementary Figure \ref{fig:convergence} for the GCAA tetraloop. The iterations with extreme melting temperature predictions (4-6) correspond to the discovery of the unfolded state. All of the Rosetta-generated structures correspond to the folded state, leading to gradual exploration of the unfolded state over the first three iterations. Iterations 4-6 learn the relative populations of the folded and unfolded states across temperature, while the final three iterations show convergence in terms of the relative populations of the folded and unfolded states.

In contrast to the GCAA tetraloop, where Rosetta only predicts the most stable structure, the Rosetta predictions for HIV-TAR constitute an ensemble (Supplementary Figure \ref{fig:hiv-tar-contours}). While TM-aMD explores new states not present in the initial ensemble for the GCAA tetraloop, HIV-TAR requires less exploration. Overall, the findings outlined in our convergence analysis show that the quality of the initial ensemble has a dramatic impact on the convergence of the TM-aMD algorithm.

\begin{figure}[h!]%
    \centering
    \includegraphics[
    width=\textwidth
    ]{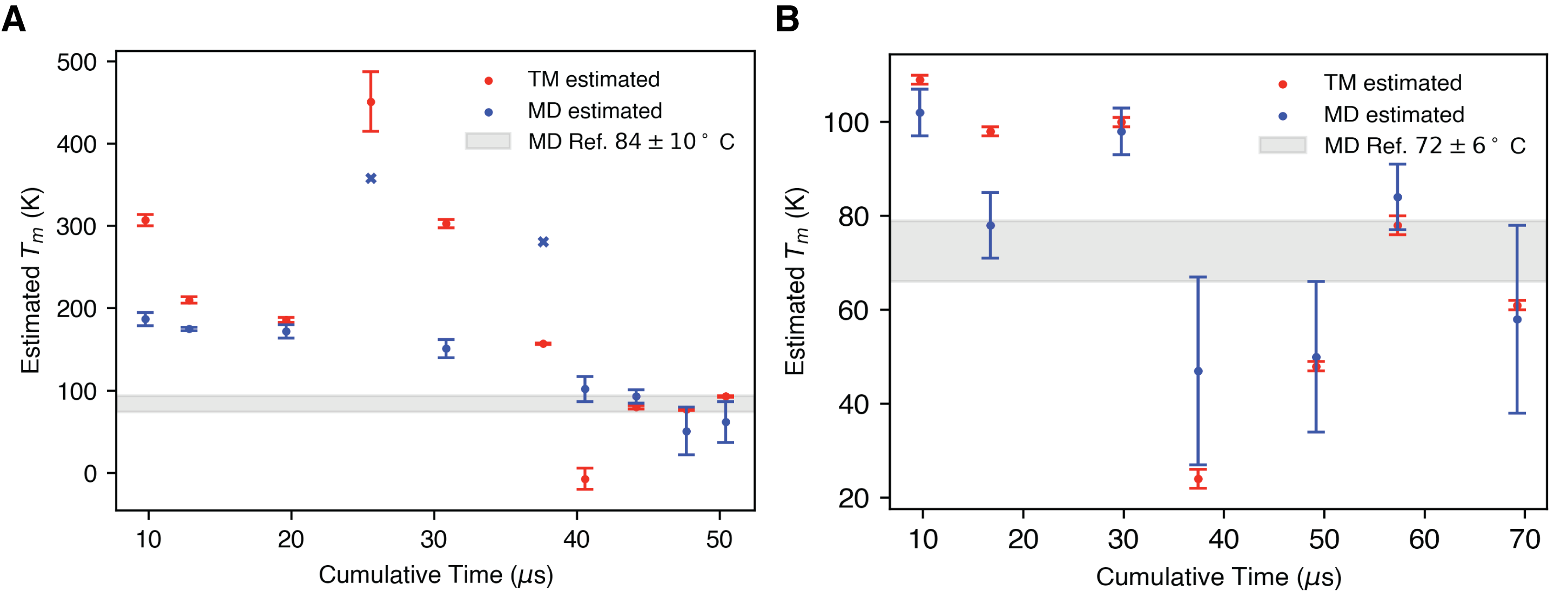}
    \caption{{\bf Convergence of Melting Temperature Calculations.} {\bf A} Melting temperatures for the GCAA tetraloop computed independently for each iteration of TM-aMD with one standard deviation uncertainties on the prediction. Convergence occurs after 45$\mu$s of MD simulation. For visual clarity, predictions with uncertainties greater than 100$^\circ$C are represented as x's. {\bf B} Melting temperatures for HIV-TAR are computed independently for each iteration of TM-aMD with one standard deviation uncertainties on the prediction. In both panels, the TM predictions show smaller uncertainties relative to the MD predictions.}
    \label{fig:convergence}%
\end{figure}

\begin{figure}[]%
    \centering
    \includegraphics[
    width=\textwidth
    ]{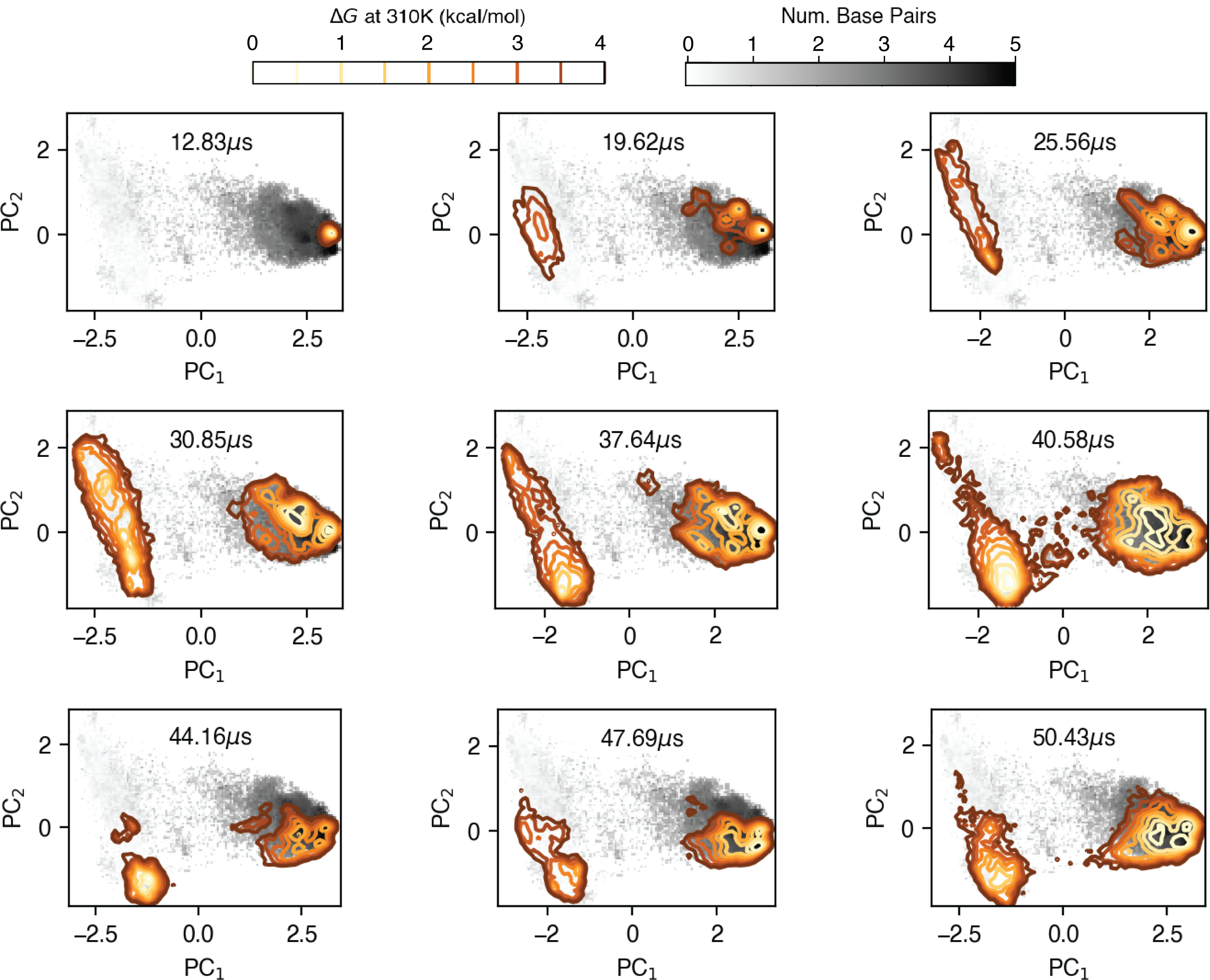}
    \caption{{\bf Convergence Towards Equilibrium for GCAA.} The TM-predicted conformational landscape at each iteration is superimposed on the distribution of all MD structures, shaded by the number of base pairs. The cumulative simulation time associated with iteration is listed.}
    \label{fig:gcaa-contours}%
\end{figure}

\begin{figure}[]%
    \centering
    \includegraphics[
    width=\textwidth
    ]{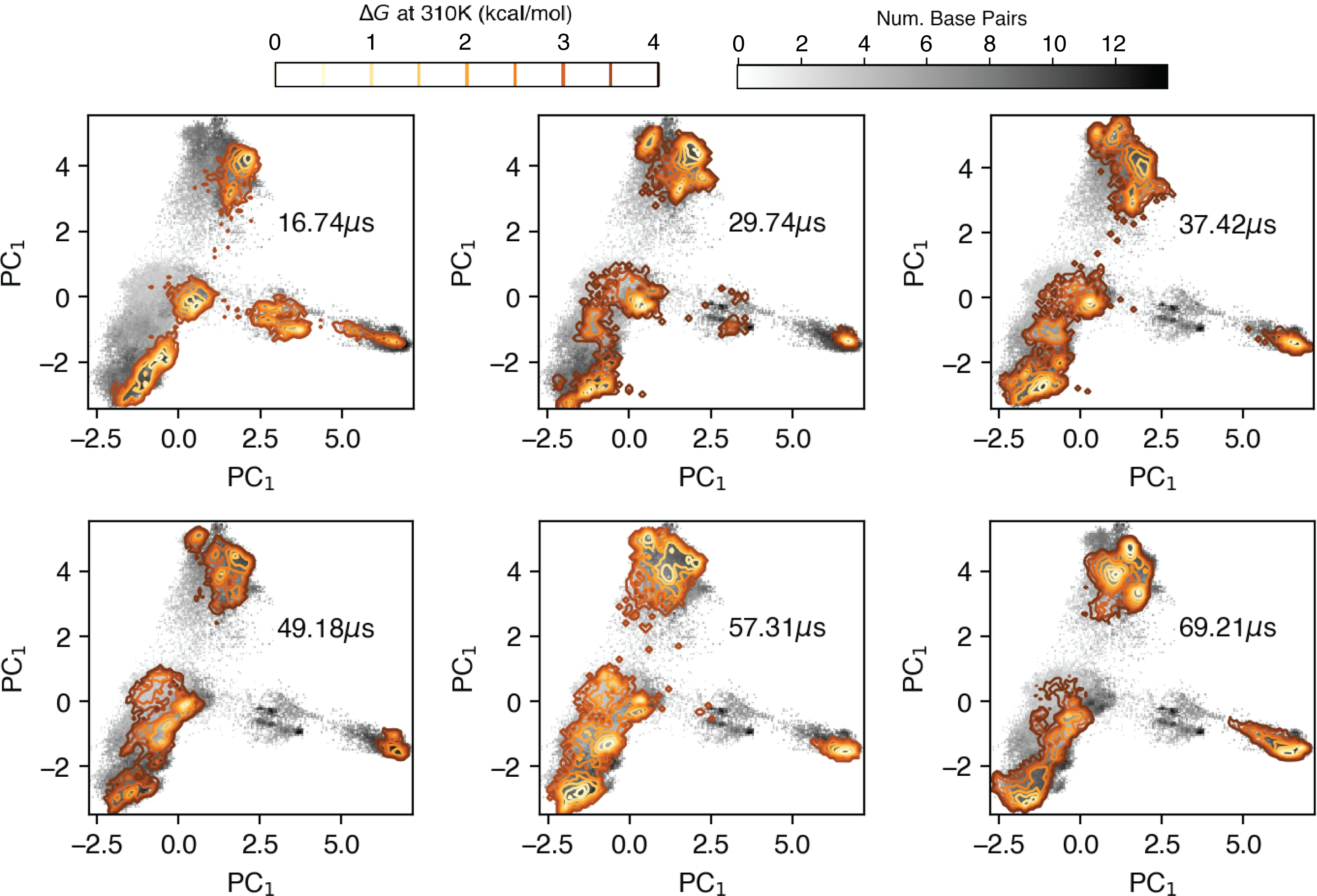}
    \caption{{\bf Convergence Towards Equilibrium for HIV-TAR.} The TM-predicted conformational landscape at each iteration is superimposed on the distribution of all MD structures, shaded by the number of base pairs. The cumulative simulation time associated with iteration is listed.}
    \label{fig:hiv-tar-contours}%
\end{figure}


\begin{thebibliography}{10}

\bibitem{wales-energy-landscapes}
D.~J. Wales,
\newblock Annual review of physical chemistry {\bf 69}, 401 (2018).

\bibitem{zwanzig-fep}
R.~W. Zwanzig,
\newblock The Journal of Chemical Physics {\bf 22}, 1420 (1954).

\bibitem{monte-carlo-review}
F.~James,
\newblock Reports on progress in Physics {\bf 43}, 1145 (1980).

\bibitem{wham}
S.~Kumar, J.~M. Rosenberg, D.~Bouzida, R.~H. Swendsen, and P.~A. Kollman,
\newblock Journal of computational chemistry {\bf 13}, 1011 (1992).

\bibitem{bar}
C.~H. Bennett,
\newblock Journal of Computational Physics {\bf 22}, 245 (1976).

\bibitem{mbar}
M.~R. Shirts and J.~D. Chodera,
\newblock The Journal of chemical physics {\bf 129} (2008).

\bibitem{parisi-replica-trick}
M.~Mezard, G.~Parisi, and M.~A. Virasoro,
\newblock {\em Spin glass theory and beyond: An Introduction to the Replica
  Method and Its Applications}, volume~9,
\newblock World Scientific Publishing Company, 1987.

\bibitem{nonequilibrium-learning}
J.~Sohl-Dickstein, E.~Weiss, N.~Maheswaranathan, and S.~Ganguli,
\newblock Deep unsupervised learning using nonequilibrium thermodynamics,
\newblock in {\em International conference on machine learning}, pages
  2256--2265, PMLR, 2015.

\bibitem{ddpm}
J.~Ho, A.~Jain, and P.~Abbeel,
\newblock Advances in neural information processing systems {\bf 33}, 6840
  (2020).

\bibitem{sgm-sde}
Y.~Song et~al.,
\newblock arXiv preprint arXiv:2011.13456  (2020).

\bibitem{ddpm-remd}
Y.~Wang, L.~Herron, and P.~Tiwary,
\newblock Proceedings of the National Academy of Sciences {\bf 119},
  e2203656119 (2022).

\bibitem{gcaa-nmr}
F.~M. Jucker, H.~A. Heus, P.~F. Yip, E.~H. Moors, and A.~Pardi,
\newblock Journal of molecular biology {\bf 264}, 968 (1996).

\bibitem{hiv-nmr}
F.~Aboul-ela, J.~Karn, and G.~Varani,
\newblock Nucleic acids research {\bf 24}, 3974 (1996).

\bibitem{excited-states}
L.~R. Ganser et~al.,
\newblock Cell reports {\bf 30}, 2472 (2020).

\bibitem{feynman}
R.~P. Feynman,
\newblock Physical Review {\bf 97}, 660 (1955).

\bibitem{TI}
A.~Gelman and X.-L. Meng,
\newblock Statistical science , 163 (1998).

\bibitem{TFEP}
C.~Jarzynski,
\newblock Physical Review E {\bf 65}, 046122 (2002).

\bibitem{NFReview}
I.~Kobyzev, S.~J. Prince, and M.~A. Brubaker,
\newblock IEEE transactions on pattern analysis and machine intelligence {\bf
  43}, 3964 (2020).

\bibitem{RealNVP}
L.~Dinh, J.~Sohl-Dickstein, and S.~Bengio,
\newblock arXiv preprint arXiv:1605.08803  (2016).

\bibitem{NeuralODE}
R.~T. Chen, Y.~Rubanova, J.~Bettencourt, and D.~K. Duvenaud,
\newblock Advances in neural information processing systems {\bf 31} (2018).

\bibitem{BG}
F.~No{\'e}, S.~Olsson, J.~K{\"o}hler, and H.~Wu,
\newblock Science {\bf 365}, eaaw1147 (2019).

\bibitem{NF-REMD}
M.~Invernizzi, A.~Kramer, C.~Clementi, and F.~No{\'e},
\newblock The Journal of Physical Chemistry Letters {\bf 13}, 11643 (2022).

\bibitem{NF-MC}
M.~Gabri{\'e}, G.~M. Rotskoff, and E.~Vanden-Eijnden,
\newblock Proceedings of the National Academy of Sciences {\bf 119},
  e2109420119 (2022).

\bibitem{bbvi}
A.~K. Dhaka et~al.,
\newblock Advances in Neural Information Processing Systems {\bf 34}, 7787
  (2021).

\bibitem{smooth-NF}
J.~K{\"o}hler, A.~Kr{\"a}mer, and F.~No{\'e},
\newblock Advances in Neural Information Processing Systems {\bf 34}, 2796
  (2021).

\bibitem{neural-spline}
C.~Durkan, A.~Bekasov, I.~Murray, and G.~Papamakarios,
\newblock Advances in neural information processing systems {\bf 32} (2019).

\bibitem{AIS-NF}
L.~I. Midgley, V.~Stimper, G.~N. Simm, B.~Sch{\"o}lkopf, and J.~M.
  Hern{\'a}ndez-Lobato,
\newblock arXiv preprint arXiv:2208.01893  (2022).

\bibitem{SNF}
H.~Wu, J.~K{\"o}hler, and F.~No{\'e},
\newblock Advances in Neural Information Processing Systems {\bf 33}, 5933
  (2020).

\bibitem{wirnsberger-TFEP}
P.~Wirnsberger et~al.,
\newblock The Journal of Chemical Physics {\bf 153} (2020).

\bibitem{LFEP}
P.~Wirnsberger et~al.,
\newblock Machine Learning: Science and Technology {\bf 3}, 025009 (2022).

\bibitem{Anderson1982}
B.~D. Anderson,
\newblock Stochastic Processes and their Applications {\bf 12}, 313 (1982).

\bibitem{onsager-solution}
S.~M. Bhattacharjee and A.~Khare,
\newblock Current science {\bf 69}, 816 (1995).

\bibitem{ising-crit-exp}
J.~Cardy,
\newblock {\em Scaling and renormalization in statistical physics}, volume~5,
\newblock Cambridge university press, 1996.

\bibitem{wolff-critical-slow-down}
U.~Wolff,
\newblock Nuclear Physics B-Proceedings Supplements {\bf 17}, 93 (1990).

\bibitem{DE-Shaw-RNA-ff}
D.~Tan, S.~Piana, R.~M. Dirks, and D.~E. Shaw,
\newblock Proceedings of the National Academy of Sciences {\bf 115}, E1346
  (2018).

\bibitem{bussi-ff-deficiency}
P.~Kuhrova et~al.,
\newblock Journal of chemical theory and computation {\bf 12}, 4534 (2016).

\bibitem{kremer-polymer}
H.-P. Hsu and K.~Kremer,
\newblock The Journal of Chemical Physics {\bf 159} (2023).

\bibitem{parisi-RNA}
A.~Pagnani, G.~Parisi, and F.~Ricci-Tersenghi,
\newblock Physical review letters {\bf 84}, 2026 (2000).

\bibitem{wales-RNA}
D.~Chakraborty, R.~Collepardo-Guevara, and D.~J. Wales,
\newblock Journal of the American Chemical Society {\bf 136}, 18052 (2014).

\bibitem{hashimi-glass}
A.~T. Frank, Q.~Zhang, H.~M. Al-Hashimi, and I.~Andricioaei,
\newblock Biophysical journal {\bf 108}, 2876 (2015).

\bibitem{kremer-ML}
A.~Banerjee, A.~Iscen, K.~Kremer, and O.~Kukharenko,
\newblock The Journal of Chemical Physics {\bf 159} (2023).

\bibitem{hashimi-invisible}
J.~Lee, E.~A. Dethoff, and H.~M. Al-Hashimi,
\newblock Proceedings of the National Academy of Sciences {\bf 111}, 9485
  (2014).

\bibitem{hashimi-nmr-dock}
A.~C. Stelzer et~al.,
\newblock Nature chemical biology {\bf 7}, 553 (2011).

\bibitem{ANNNI}
M.~E. Fisher and W.~Selke,
\newblock Physical Review Letters {\bf 44}, 1502 (1980).

\bibitem{hopfield-network}
J.~J. Hopfield,
\newblock Proceedings of the national academy of sciences {\bf 79}, 2554
  (1982).

\bibitem{amit-spin-glass}
D.~J. Amit, H.~Gutfreund, and H.~Sompolinsky,
\newblock Physical Review Letters {\bf 55}, 1530 (1985).

\bibitem{how-to-think}
Q.~Vicens and J.~S. Kieft,
\newblock Proceedings of the National Academy of Sciences {\bf 119},
  e2112677119 (2022).

\bibitem{metadynamics-review}
O.~Valsson, P.~Tiwary, and M.~Parrinello,
\newblock Annual review of physical chemistry {\bf 67}, 159 (2016).

\bibitem{rosetta}
R.~Das and D.~Baker,
\newblock Annu. Rev. Biochem. {\bf 77}, 363 (2008).

\bibitem{kresten-jacs}
A.~Oxenfarth et~al.,
\newblock Journal of the American Chemical Society  (2023).

\bibitem{debenedetti-rna}
G.~H. Zerze, P.~M. Piaggi, and P.~G. Debenedetti,
\newblock The Journal of Physical Chemistry B {\bf 125}, 13685 (2021).

\bibitem{gcaa-melt}
J.~P. Sheehy, A.~R. Davis, and B.~M. Znosko,
\newblock Rna {\bf 16}, 417 (2010).

\bibitem{gvecs-bussi}
S.~Bottaro, F.~Di~Palma, and G.~Bussi,
\newblock Nucleic acids research {\bf 42}, 13306 (2014).

\bibitem{hiv-tar-binding}
S.~S. Chavali, S.~M. Mali, J.~L. Jenkins, R.~Fasan, and J.~E. Wedekind,
\newblock Journal of Biological Chemistry {\bf 295}, 16470 (2020).

\bibitem{hiv-experiment}
A.~L. Smith, J.~Kassman, K.~J. Srour, and A.~M. Soto,
\newblock Biochemistry {\bf 50}, 9434 (2011).

\bibitem{dill-chen}
S.-J. Chen and K.~A. Dill,
\newblock Proceedings of the National Academy of Sciences {\bf 97}, 646 (2000).

\bibitem{folding-at-home}
A.~L. Beberg, D.~L. Ensign, G.~Jayachandran, S.~Khaliq, and V.~S. Pande,
\newblock Folding@ home: Lessons from eight years of volunteer distributed
  computing,
\newblock in {\em 2009 IEEE International Symposium on Parallel \& Distributed
  Processing}, pages 1--8, IEEE, 2009.

\bibitem{FAST}
M.~I. Zimmerman and G.~R. Bowman,
\newblock Journal of chemical theory and computation {\bf 11}, 5747 (2015).

\bibitem{REAP}
Z.~Shamsi, K.~J. Cheng, and D.~Shukla,
\newblock The Journal of Physical Chemistry B {\bf 122}, 8386 (2018).

\bibitem{unet}
O.~Ronneberger, P.~Fischer, and T.~Brox,
\newblock U-net: Convolutional networks for biomedical image segmentation,
\newblock in {\em Medical Image Computing and Computer-Assisted
  Intervention--MICCAI 2015: 18th International Conference, Munich, Germany,
  October 5-9, 2015, Proceedings, Part III 18}, pages 234--241, Springer, 2015.

\bibitem{adam}
D.~P. Kingma and J.~Ba,
\newblock arXiv preprint arXiv:1412.6980  (2014).

\bibitem{ddpm-self-conditioning}
T.~Chen, R.~Zhang, and G.~Hinton,
\newblock arXiv preprint arXiv:2208.04202  (2022).

\bibitem{equivariant-diffusion-model}
E.~Hoogeboom, V.~G. Satorras, C.~Vignac, and M.~Welling,
\newblock Equivariant diffusion for molecule generation in 3d,
\newblock in {\em International conference on machine learning}, pages
  8867--8887, PMLR, 2022.

\bibitem{three-interpretations}
C.~Luo,
\newblock arXiv preprint arXiv:2208.11970  (2022).

\bibitem{ddim}
J.~Song, C.~Meng, and S.~Ermon,
\newblock arXiv preprint arXiv:2010.02502  (2020).

\bibitem{openmm}
P.~Eastman et~al.,
\newblock PLoS computational biology {\bf 13}, e1005659 (2017).

\bibitem{tip4p}
J.~L. Abascal and C.~Vega,
\newblock The Journal of chemical physics {\bf 123} (2005).

\bibitem{tram}
H.~Wu, F.~Paul, C.~Wehmeyer, and F.~No{\'e},
\newblock Proceedings of the National Academy of Sciences {\bf 113}, E3221
  (2016).

\bibitem{landau-lifshitz}
L.~D. Landau and E.~M. Lifshitz,
\newblock {\em Statistical Physics: Volume 5}, volume~5,
\newblock Elsevier, 2013.

\bibitem{msm}
S.~Bacallado, J.~D. Chodera, and V.~Pande,
\newblock The Journal of chemical physics {\bf 131} (2009).

\end{thebibliography}
\end{document}